
%
%
\magnification=\magstep1
\advance\voffset15mm
\advance\mathsurround1pt
\newskip\diaskip \diaskip=9mm
\let\cl\centerline
\def\ref#1{\hbox{${}^{#1)}$}}
\def\gev{\hbox{\rm  GeV}}
\def\intdppi{\int\!{d^4p\over(2\pi)^4}\,\,}
\def\intdp{\int\!d^4p\,\,\,}

\def\pimm{{\Pi}_\mu^{\!\phantom{\mu}\mu}(0)}

\def\Tr{\mathop{\hbox{\rm Tr}}}          
\def\CKM{Cabibbo-Kobayashi-Maskawa}
\def\0{(0)}
\def\2{^2}
\def\half{1/2}

\let\phi\varphi



\def\mwsq{\hbox{$m_{\rm W}^{\phantom2}{}^{\!2}$}}
\def\mzsq{m_{\rm Z}^{\hphantom2}{}^{\!2}}

\def\mtopsq{\hbox{$m_{\rm t}^{\,\,2}$}}

\def\ksq{k^2}
\def\psq{p^2}
\def\phat{\hbox{$\not\mkern-3.3mu p\mkern3.3mu$}}

\def\at{a^2}
\def\cv{c_v}
\def\ca{c_a}
\def\gwsq{g_w{}^2}
\def\thw{\theta_{\!w}}
\def\cwsq{\cos^{\,2}\!\thw}
\def\swsq{\sin^{\,2}\!\thw}
\def\Gf{G_{\!F}^{\phantom2}}
\def\gmns{g^{\mu\nu}}
\def\gmn{g_{\mu\nu}}
\def\gfive{\gamma_5}
\def\cgamma#1{(\cv\gamma_{#1}-\ca\gamma_{#1}\gfive)}
\def\C{C}
\def\Feynman{\hfill Feynman diagram to be added\quad}
%
\def\qed                                {1}
\def\gf                                 {2}
\def\tract  {3}
\def\bosons {4}
\def\carter {5}
\def\sir    {6}
\def\dona   {7}
\def\data   {8}

\font\ninerm = cmr9
\headline={\it J. J. Lodder, Preprint, \hfill
hep-th/9405264
\hfill
Submitted to Physics Letters \thinspace {\bf B},
1994, April 7}

\cl{\bf\uppercase{Recalculation
of W and Z radiative corrections}}
\smallskip
\cl{\bf\uppercase{and a top quark mass estimate}}
\bigskip
\cl{ J. J. LODDER}
\smallskip
\cl{\it Oudegracht 331\thinspace bis,
3511\thinspace\thinspace PC Utrecht,
The Netherlands}
\medskip
\cl{H. J. de BLANK\footnote{${}^\dagger$}{Present address:
Max-Planck Institut f\"ur Plasmaphysik, EURATOM-IPP
Association, 85740 Garching, FRG.}}
\smallskip
\cl{\it Stuntzstra\ss e 37, 81677 Munich, FRG }
\bigskip\bigskip\bigskip
{\ninerm
The newly published top mass prediction of 85~\gev\
conflicts with indirect limits on the top mass,
based on precision measurements interpreted by
calculations using dimensional regularization.
The top quark contribution to part of the electroweak W and
Z vacuum polarization tensor is recomputed using the
symmetrical theory of generalised functions.
The squared top mass coefficient is larger  by a
factor 1.44.
The indirect limits on the top mass based on
this computation should therefore be divided by 1.2.
Further recomputation and data analysis is needed.}

\bigskip
\noindent{\bf 1. Introduction}
\bigskip
It has been shown\ref\qed{} that the symmetrical theory of
generalised functions\ref{\gf,\tract} can be applied to
computations in quantum field theory.
All results are automatically finite and regularization is
never needed to get rid of infinities.
Integrals which are infinite in the standard sense may or
may not be determinate in the generalised sense, depending
on the scale transformation properties of the result.
A determinacy calculus,
using an indeterminate constant~$\C$,
can be set up for book-keeping of the determinacy\ref\tract.
Only determinate results can have physical meaning.

Application to QED, assuming gauge invariance and
completeness of the standard model,
leads to a prediction\ref\bosons{} of~\hbox{$85.1
\pm0.3$}~\gev\ for the mass of the top quark.

Indirect limits on the top quark mass have been obtained
by comparing computed electroweak radiative corrections
to the outcome of precision experiments\ref\carter.
The results for the top mass are substantially higher
than the 85.1~\gev\ prediction.
However,
the interpretation of the experiments has been based on
theoretical formul\ae\ derived
by means of dimensional regularization.
The symmetrical theory of generalised functions applied to
the same diagrams gives results
which differ by finite renormalizations.

Upon computation of the simplest term it is found that the
top mass estimate is lowered.
The final result to second order (one loop) is a reduction
of the top mass estimate by a factor
of~$0.84$.
This removes at least part of the conflict
between the~85.1~\gev\ prediction and the
indirect
experimental data, as commonly interpreted.

On basis of dimensional regularization it is not
possible to predict the top mass.
It is therefore inconsistent  to confront the top mass
prediction\ref\bosons{} with experimental interpretations
depending essentially on dimensional regularization.

It is not sufficiently appreciated that all regularization
methods are arbitrary by finite renormalizations.
These can be fixed by convention,
but there is no reason to suppose one convention to be
superior to others.
Results depending on arbitrary conventions cannot serve as
basis for physical predictions.
The reasons for preferring a form of dimensional
regularization are irrelevant when gauge invariance
has been secured by other means\ref\qed.
Observability of a physical quantity does not imply that
its calculation is independent of the regularization
method.

The symmetrical theory of generalised functions,
applied to computations in quantum field theory,
is not arbitrary by finite renormalizations.
The arbitrariness of the product definition
is resolved by imposing simplicity arguments
on the underlying mathematical theory,
and the values of integrals follow uniquely from the product
definition.

Our preliminary conclusion is that the experimental
data, interpreted on basis of dimensional
regularization,
cannot be used to reject the~85.1~\gev\ top mass prediction.
It is not possible at present to be more explicit,
since it is not clear from the literature how the data
reduction has been done and in how far the results depend
on the use of theoretical formul\ae\ in which the use of
dimensional regularization does make a difference.

Other calculations and some of the data reduction should be
redone as well before a firm conclusion can be reached.
This entails much work, so the top
quark (if it is near 85~\gev) might  be found before
this can be accomplished.

\bigskip
\noindent{\bf 2. W and Z vacuum polarization by fermions}
\bigskip\noindent
In this letter we compute the fermionic contributions to
the vacuum polarization,
in particular the correction to the parameter~$\rho$
defined to zeroth order by\ref{\sir,\dona}
$$\rho:={\mwsq\over\cwsq\,\mzsq}.\eqno(1)$$
In general the vacuum polarization
tensor~$\Pi^{\mu\nu}(k)$ can be written as
$$\Pi^{\mu\nu}(k)=
A(\ksq)\gmns+B(\ksq)k^\mu k^\nu.\eqno(2)$$
The correction to~$\rho$ is then found to
be
\ref{\sir\dona}{}
$$\Delta\rho=
{A_{\rm ZZ}\0\over\mzsq}-{A_{\rm WW}\0\over\mwsq},\eqno(3)$$
where the subscripts indicate the gauge boson type.

The fermionic  contribution to the vacuum polarization is
found from the fundamental two fermion-boson vertex and the
corresponding loop diagram
\vskip\diaskip
\vbox to 0pt{\vss\line{\Feynman \hfill(4)}}
\vskip\diaskip
\noindent
which is needed for this letter only at boson
momentum~$k=0$. Here~$g$, $\cv$, and~$\ca$ are coupling
constants. Substitution of the Feynman rules for Dirac
fermions gives
$$\Pi_{\mu\nu}(0)=
-g^2\Tr\intdppi\,\cgamma\mu
{\phat+m_1\over p^2-m_1{}^2}\cgamma\nu
{\phat+m_2\over p^2-m_2{}^2}.
\eqno{(5)}$$
After evaluating the trace it is convenient to contract
with~$g^{\mu\nu}$ to obtain
$$4iA(0)=\pimm=2g^2
\intdp{(\ca^2+\cv^2)p^2+2(\ca^2-\cv^2)m_1m_2\over
(p^2-m_1{}^2)(\psq-m_2{}^2)}.
\eqno{(6)}$$
For W bosons the dominant contribution is obtained by
taking the fermions to be a top and a bottom quark.
To sufficient accuracy the bottom quark is massless.
The \CKM\ matrix element then equals one.
(Considering all massless quark flav\-ours
the CKM matrix elements sum to one by unitarity,
giving the same result).
Combining the denominators by the standard Feynman trick,
and substituting the values
of the integrals from the appendix,
yields after evaluation of the trivial auxiliary integration
$$A_{\rm WW}(0)=
{3\gwsq\mtopsq\over32\pi^2}(1-\log\mtopsq-\C),
\eqno{(7)}$$
with an additional factor~$3$
to account for the colours of the quarks.
It is seen from the appearance of a~$\C$
that the result is indeterminate.

For the Z boson the dominant correction
comes from the 
top-antitop loop,
which can be evaluated directly.
The result is
$$A_{\rm ZZ}(0)={3\gwsq\mtopsq\over32\pi^2\cwsq}
\big({1\over2}-{4\over3}\sin^2\thw
+{16\over9}\sin^4\thw-\log\mtopsq -
\C\big).\eqno{(8)}$$
In contrast to the corresponding determinate
photon mass correction\ref\gf{}
which has only~$\cv$ vector coupling,
the axial~$\ca$ contribution is
indeterminate.
The corrections
for the members of the W triplet are not
independent.
Substituting~(7) and~(8) into~(3),
the correction
produced by the top quark
to the ratio~$\rho$ defined in~(1)
is seen to be determinate
$$\Delta\rho_t={3\sqrt2\over16\pi^2}\Gf\mtopsq
(1+{8\over3}\sin^2\thw-{32\over9}\sin^4\thw).\eqno{(9)}$$
This is no surprise since the corresponding
computation\ref{\sir,\dona}  using dimensional
regularization yields a finite answer
which must agree with~(9) up to a finite
renormalization\ref\qed.
The result has been expressed
in terms of the Fermi coupling
constant~$\Gf=\gwsq/4\sqrt2\,\mwsq$

Comparing the result
using symmetrical generalised functions
to the result of the same calculation
using dimensional regularization\ref{\sir,\dona}
$$\Delta\rho_t={3\sqrt2\over16\pi^2}\Gf\mtopsq,\eqno{(10)}$$
one sees that (using~$\sin^2\thw=0.2325$\ref\data) the
coefficient of the squared top mass increases by a factor
of~$1.44$,
so all else remaining the same the top mass limits
predicted on basis of (10)
must be divided by~$1.2$.
Given the large uncertainties inherent in the experimental
data this removes part (and perhaps all) of the
discrepancy of the~$85.1$~\gev\ prediction and indirect
estimates ranging from~100-150~\gev.
Conversely,
if the top mass prediction from\ref\bosons{}
$$\mtopsq=9\mwsq/8,\eqno{(11)}$$
is substituted the prediction
$$\Delta\rho_{t}={27\gwsq\over512\pi^2}
(1+{8\over3}\sin^2\thw-{32\over9}\sin^4\thw)
\approx0.0023\cdot1.44=0.0033,\eqno{(12)}$$
is obtained.
This cannot yet be compared with experiment since the other
contributions to~$\Delta\rho$ have to be recomputed as well.
It is unclear at present how this affects the
interpretation of experiments.
(Precision measurements of~$\swsq$
often give several results
for different top mass assumptions,
but the coefficient multiplying~$\mtopsq$ may have to be
recomputed as well).
This will affect all top mass coefficients from
quadratically divergent diagrams.

\bigskip
\noindent{\bf 3 Conclusion}
\bigskip
It is inconsistent to compare the top quark mass
prediction\ref\bosons{} with indirect
limits on the top mass based on computations using
dimensional regularization.
The results obtained by
computing with generalised functions differ by finite
renormalizations from those obtained by means of
dimensional regularization.

Recomputation of the W and Z vacuum
polarization tensor to one loop removes at least part of
the conflict between the~85.1~\gev\ top mass prediction
and experimental data interpreted on basis of dimensional
regularization. Further recomputation and data analysis is
needed to obtain definitive results.

\bigskip\noindent
{\bf Appendix}\quad
The imaginary parts of the integrals one needs were derived
in ref~\gf.
The results are
(correcting for the modified sign of~$\gmn$)
$$\eqalignno{\intdp{1\over(\psq-\at)^2}&=
-i\pi^2(\log\at+\C),&A1\cr
\intdp{\psq\over(\psq-a^2)^2}&=
-2i\pi^2\at(\log\at+\C-{1\over2})&A2\cr}$$
with~$\C$ the indeterminate constant
and the~$i\epsilon$ in the denominator
understood.
Purists may read~$\log\at$ as~$\log\at/M^2$,
with~$M$ an arbitrary unit of mass,
but by definition\ref\tract{}
this does not influence any physical result.

\noindent
It is conventional to normalize the coupling constants as
\tabskip=0ptplus1fill
$$\halign to \hsize{
#&$#$&\hbox{$#$}&$#$\cr
\hbox{boson}&g^2&\cv&\ca\cr
W&g_w{}^2/8&1&-1\cr
Z&g_w{}^2/4\cwsq&\half-4/3\sin^2\thw&\half\cr}$$
The CKM element in the W-coupling~$U_{tb}=1$
for our computation, as explained above.
\vfill\eject
\noindent
{\bf Note added 18-05-94:}
\smallskip
\noindent
If recent top mass claims near 170 \gev\ are
confirmed, the mass sum rule derived in ref.4 indicates
the need for adding charged (Higgs?) bosons to the
standard model, with a mass given by
$$m_H^2 = {8\over3} m_t^2 - 3 m_W^2$$ which
yields~$m_H=205$ \gev\ for 170 \gev\ top quarks. (assuming
again that no additional charged fermions and bosons
exist). The possible observation of heavy top quarks does
not affect the conclusions of this letter, but it will
affect predicted Higgs boson masses.
\bigskip

\noindent{\bf References}
\medskip
\parindent=-1.5em
\leftskip=1.5em
\def\ref#1{\par\leavevmode\rlap{#1)}\kern1.5em}
\def\dummyref{\par\leavevmode\kern1.5em}
\def\aut#1{{\rm#1}}                     \let\author        =\aut
\def\tit#1{{\sl#1}}                     \let\title          =\tit
\def\yea#1{{\rm(#1)}}   \let\year            =\yea
\def\pub#1{{\rm#1}}                     
\def\jou#1{{\rm#1}}                     
\def\vol#1{{\bf#1}}                     
\def\pag#1{{\rm#1}}                     \let\page            =\pag
                     \let\place          =\pla

\let\lin\relax

\ref{1}
 \aut{Lodder J.J.},
 \lin\jou{Physica}
 \vol{120A}
 \yea{1983} \pag{1, 30, 566, 579}.

\ref{2}
 \aut{Lodder J.J.},
 \lin\jou{Physica}
 \vol{116A}
 \yea{1982} \pag{45, 59, 380, 392},
 \jou{Physica}
 \vol{132A}
 \yea{1985} \pag{318}.

\ref{3}
 \aut{Lodder J.J.},
 \tit{Towards a Symmetrical Theory
     of Generalised Functions},
 \jou{CWI tract }\vol{79}
 \yea{1991},
 \pub{CWI, Amsterdam}.

\ref{4}
 \aut{Lodder J.J.},
 \tit{Quantum Electrodynamics Without Renormalization
 Part V, Bosonic contributions to the photon mass},
 \lin\jou{Submitted to Physica}
 \vol{A}
\yea{1994}.

\ref5
\aut{Carter J. R.},
\vol{Proc. LP-HEP 91, Geneva}
\year{1992},
\page{29},
\pub{World Scientific},
\place{Singapore}

\ref6
\aut{Sirlin A.},
\jou{Phys. Rev.}
\vol{D22}
\yea{1980}
\page{971},
\aut{Marciano W. J., Sirlin A.},
\jou{Phys. Rev.}
\vol{D29}
\yea{1984}
\page{945}.

\ref{7}
\aut{Donoghue, J. F., Golowich E., Holstein B. R.},
\tit{Dynamics of the standard model},
\pub{Cambridge University Press},
\yea{1992}.

\ref8
 \aut{Particle Data Group},
 \jou{Phys. Rev.}
 \vol{D45}
 \yea{1992}
 \page{part 2.}

\bye